\documentclass[10pt]{iopart}
\newcommand{\eqref}[1]{Eq.~\ref{#1}}
\usepackage{graphicx}
\usepackage{amssymb}
\usepackage{hyperref}
\usepackage{braket}

\usepackage{xspace} %

\usepackage{amsfonts}
\usepackage[figuresright]{rotating}  
\usepackage{psfrag}
\usepackage{subfigure}
\usepackage{multirow}
\usepackage{tabularx}
\usepackage{textcomp}

\usepackage{verbatim}%
\usepackage[colorinlistoftodos, textwidth=4cm, shadow]{todonotes}

\newcommand{\beginsupplement}{%
        \setcounter{table}{0}
        \renewcommand{\thetable}{S\arabic{table}}%
        \setcounter{figure}{0}
        \renewcommand{\thefigure}{S\arabic{figure}}%
     }

\def\size{0.85}

\def\nt{\tilde{n}}
\def\beqa{\begin{eqnarray}}
\def\eeqa{\end{eqnarray}}
\def\beq{\begin{equation}}
\def\eeq{\end{equation}}

\def\psie{\hat{\mathcal{E}}}
\def\psip{\hat{\mathcal{P}}}
\def\psis{\hat{\mathcal{S}}}

\def\sRp{w_0} %

\def\Veff{V^{\rs eff}}
\def\psieWf{{\mathcal{E}}}

\def\up{\uparrow}
\def\down{\downarrow}

\hypersetup{
    colorlinks=false,
    pdfborder={0 0 0},
}

\newcommand{\rs}{\rm \scriptscriptstyle}
\def\D{\ensuremath{\Delta}\xspace}
\newcommand{\figref}[1]{Fig.~\ref{#1}}

\def\s{\ensuremath{\mathcal{S}}\xspace}
\def\nn{\nonumber}
\def\s{\ensuremath{\sigma}\xspace}
\def\a{\ensuremath{\alpha}\xspace}

\def\d{\ensuremath{\delta}\xspace}
\def\D{\ensuremath{\Delta}\xspace}
\def\om{\ensuremath{\omega}\xspace}
\def\Om{\ensuremath{\Omega}\xspace}

\newcommand{\ra}{\ensuremath{\rightarrow}\xspace}
\def\bR{{\bf R}}
\def\br{{\bf r}}
\def\bz{{\bf z}}
\def\counterpropagating{counter-propagating\xspace}

\renewcommand{\braket}[2]{\left\langle #1\middle| #2\right\rangle}
\newcommand{\integral}[1]{\int \! \mathrm{d} #1\,}                    %
\newcommand{\integralb}[3]{\int\limits_{#1}^{#2} \! \mathrm{d} #3\,}  %

\newcommand\sbullet[1][.5]{\mathbin{\vcenter{\hbox{\scalebox{#1}{$\bullet$}}}}}

\usepackage{color}
\usepackage[normalem]{ulem}

\begin{document}
\title{Two photon conditional phase gate based on Rydberg slow light polaritons}

\author{Przemyslaw~Bienias}
\address{ Joint Quantum Institute, NIST/University of Maryland, College Park, Maryland 20742, USA}
\address{Joint Center for Quantum Information and Computer Science, NIST/University of Maryland, College Park, Maryland 20742, USA}
\address{Institute for Theoretical Physics III, University of Stuttgart, Germany}
\ead{bienias@umd.edu}

\author{Hans~Peter~B\"{u}chler}
\address{Institute for Theoretical Physics III and Center for Integrated Quantum Science and Technology, University of Stuttgart, 70550 Stuttgart, Germany}

\date{\today}

\date{\today}

\begin{abstract}
We analyze the fidelity of a deterministic quantum phase gate for  two photons  counterpropagating as polaritons through a cloud of Rydberg atoms under the condition of electromagnetically induced transparency (EIT).
We provide analytical results for the  phase shift of the quantum gate, and provide an estimation for all processes leading to a reduction to the gate fidelity. Especially, the influence of
 losses form the intermediate level,  dispersion of the photon wave packet,   scattering into additional polariton channels, finite lifetime of the Rydberg state, as well as effects of transverse size of the wave packets are accounted for. We show that  the flatness of the effective interaction, caused by  the blockade phenomena, suppresses the corrections due to the finite transversal size. This is a strength of Rydberg-EIT setup compared to other approaches.
Finally, we provide the experimental requirements for the realization of a high fidelity quantum phase gate using Rydberg polaritons. 
\end{abstract}
\maketitle
\ioptwocol
{\it Introduction ---}
Photons  interact extremely weakly with each other, propagate with the speed of light and provide a high bandwidth. These three features make photons an excellent carrier of %
information.
However, for applications in quantum %
information processing \cite{Nielsen2000}, interactions on the level of single quanta are necessary. %
Such interactions can be achieved by coupling photons to %
matter \cite{Turchette1995,Fushman2008,Parigi2012,Volz2014,Reiserer2014}. Especially,  Rydberg-EIT (rEIT) has emerged a promising approach \cite{Fleischhauer2005,Lo2011,Shiau2011,Feizpour2015,Nemoto2004,Mohapatra2007,Pritchard2010} visible in  great experimental 
 \cite{Mohapatra2008,Parigi2012,Peyronel2012,Dudin2012,Dudin2012b,Maxwell2013, Hofmann2013,Firstenberg2013,Gorniaczyk2014,Baur2014,Tiarks2014,Tresp2015, Gorniaczyk2016,Tresp2016,Tiarks2016,Schine2016} and
  theoretical~\cite{Sevincli2011,Gorshkov2011,Gorshkov2011,Petrosyan2011, Otterbach2013,Gorshkov2013,Stanojevic2013,Liu2014,He2014a,Grankin2014,Li2014,Wu2014,Bienias2014,Sommer2015,Grankin2015, Lin2015,Moos2015,Maghrebi2015,Maghrebi2015c,Caneva2015,Shi2015,Jachymski2016,Gullans2016,Murray2016b,Zeuthen2016,Li2015a,Murray2017,Bienias2016}
progress.
A photonic quantum phase gate using Rydberg-EIT in a \counterpropagating setup 
was first discussed by Friedler et~al.~\cite{Friedler2005} 
and an extended description was shown by Gorshkov et~al.~\cite{Gorshkov2011}.
However, a study including all effects that decrease the fidelity of the phase gate, as well as, a discussion of a specific microscopic setup is still missing. Here, we attempt to fill this gap by analyzing within a microscopic description the different source for a reduction in gate fidelity in a realistic setup.

Recently, a few alternative approaches for a quantum phase gate have emerged  \cite{Paredes-Barato2014a,Khazali2015}, which were motivated 
 (at least partially) by the belief that %
the link between propagation and interaction \cite{Shapiro2006,Shapiro2007} precludes high-fidelity gates whenever a cross-phase modulation (XPM) on a single photon level is used. 
However, %
for the Rydberg-EIT setup the interaction between Rydberg polaritons has a finite range and Rydberg polaritons acquire an effective mass in addition to the linear slow light velocity, and therefore the proposed no-go theorems~\cite{Shapiro2006,Shapiro2007,Gea-Banacloche2010} do not apply \cite{Bienias2016}.

The  effective interaction between Rydberg polaritons has an important feature, namely that the interaction potential
 is flat~\cite{Bienias2014} for distances smaller than
interaction range (called blockade radius which is on the order of $10\mu$m). %
This feature could in principle enable a phase gate in a copropagating configuration, which in general  is characterized by longer interaction times and therefore could enable greater phase shifts.
However, the condition of a homogeneous phase shift requires  compression of photons to the size of the blockade radius. 
This violates  the conditions for neglecting the mass term which depicts the quadratic corrections to the polaritonic dispersion relation. 
In turn, adding the mass term leads to strong mode distortion precluding the realization of a photonic quantum gate in a copropagating setup.
Thus, the phase gate might be possible only if photons pass each other  by for example being \counterpropagating \cite{Masalas2004,Andre2005,Friedler2005,Shahmoon2011} or having different group velocities  \cite{Marzlin2010}.

In this manuscript, we study a quantum phase gate for photons in a \counterpropagating setup using the microscopic theory describing the Rydberg slow light polaritons. 
The main focus is on providing an analytical expression for the imprinted phase, as well as a detailed analysis of different sources for a reduction in the gate fidelity. In addition to standard effects such as losses from the intermediate state and Rydberg levels, we also include propagation effects such a dispersion of the wave packet and energy dependence 
of the phase shift, as well as the influence of scattering into different polariton channels. Note, that our approach does not rely on semiclassical approximations recently applied  \cite{He2014a}, but is based on the full microscopic quantum description for two photons. We provide the experimental requirements for the realization of a high fidelity quantum phase gate using Rydberg polaritons.

\begin{figure}[]
\includegraphics[width=1\columnwidth]{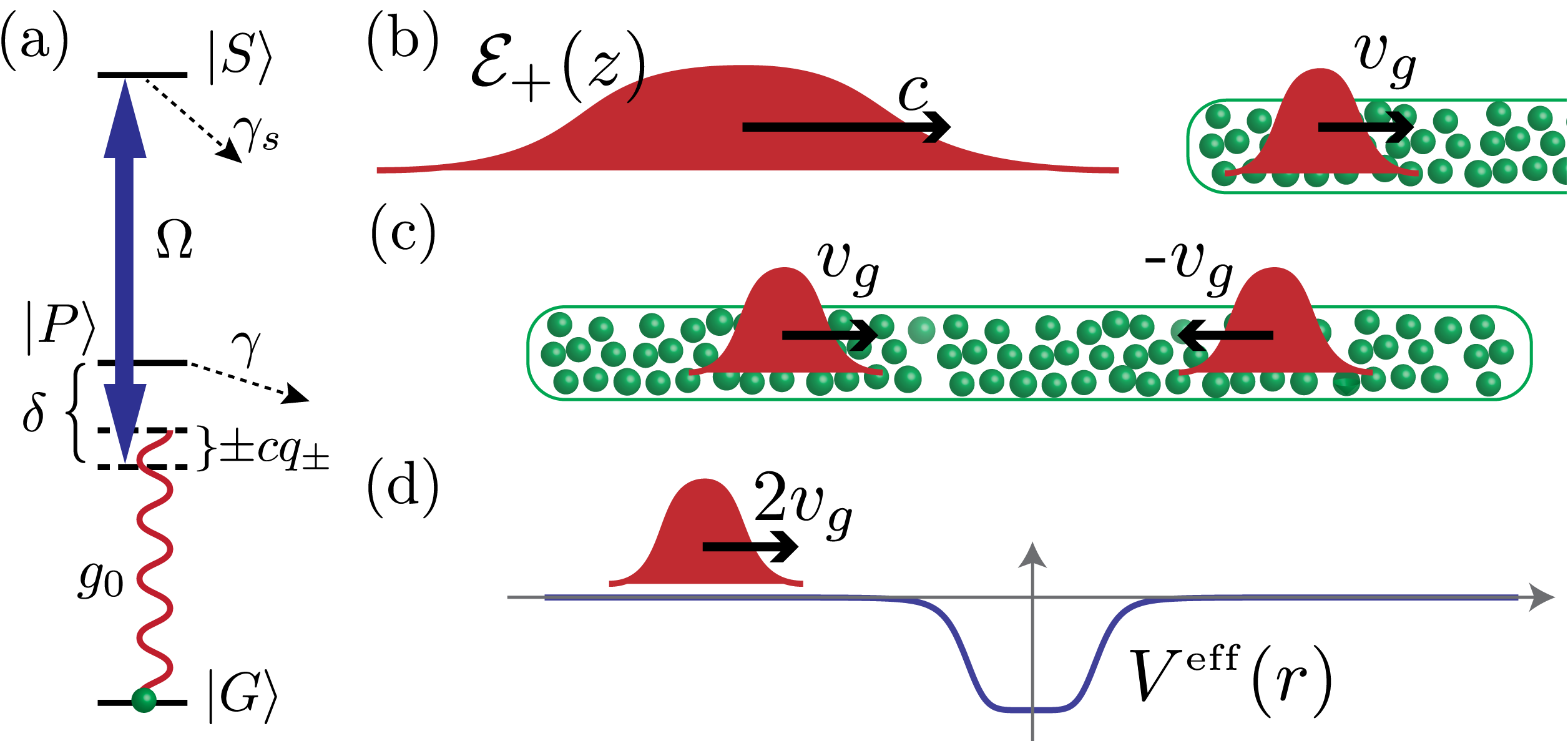}
\caption{
(a)~The probe field couples the atomic ground state $|G\rangle$ to the $p$-level $|P\rangle$
with the single-particle coupling strength $g_{0}$, while a strong coupling laser drives the transition between the $p$-level and the Rydberg
state $|S\rangle$ with Rabi frequency $2\Omega$ and detuning $\delta$.
Furthermore, $  \gamma$  denotes the halfwidth of the $p$-state and $\gamma_s$ the halfwidth the $s$-state. 
The single-particle coupling $g_0$ is related to the collective coupling $g = \sqrt{n_{\rs at}} g_0$ with $n_{\rs at}$ the particle density. 
Note that the kinetic energy of the photons $\pm\hbar c q_\pm$ accounts for the difference to the EIT condition, and in the position space takes the form $\mp i \hbar c \partial_{z}$. 
 (b)~Photon entering the EIT medium %
 is compressed by factor $v_g/c$. %
 For pulses shorter than the length of the medium, photons enter the medium without backscattering and losses because interaction between polaritons can be neglected and spatial  density variations are smooth on the scales of photonic wavelength.
 (c)~Two \counterpropagating polaritons inside the medium. (d) In the center of mass frame, the problem simplifies to the scattering on the effective potential~$V^{\rs eff} (r)$, which leads to the interaction induced phase shift. 
}
\label{fig1}
\end{figure}

{\it General concept}---
The qubit is encoded in the polarization of the photons, and  the excitation into the Rydberg levels depends on this polarization %
\cite{Firstenberg2013}, Then, photons within the Rydberg medium are propagating as Rydberg polaritons, see \figref{fig1}. 
After passing each other, they pick-up the phase shift due to the strong interaction between Rydberg atoms, 
leading to the  interaction induced phase shift only between photons having a specific polarization. We will discuss a microscopic setup  in detail at the end of the manuscript.
We define the fidelity $F$ and phase shift $\phi$ of the gate using %
 the overlap between the two-photon wave-functions with ($\psieWf\psieWf$) and without (${\psieWf\psieWf^{V=0}}$) the interaction $V$ between the polaritons: 
\begin{equation}
\sqrt{F}e^{i\phi} = 
{
\braket{\psieWf\psieWf^{V=0}}{\psieWf\psieWf}
}
.
\label{fidelityDef}
\end{equation}
Our definition of the fidelity includes all detrimental effects and is more restrictive than commonly used~\cite{He2012,He2014a} conditional-fidelity  in which two output states are normalized~\cite{SuppBienias},
\begin{equation} \sqrt{F_{\rs cond}}e^{i\phi} = \braket{\psieWf\psieWf^{V=0}}{\psieWf\psieWf}/\braket{\psieWf\psieWf^{V=0}}{\psieWf\psieWf^{V=0}}
.
\label{fidelityDef}
\end{equation}
{
We will identify the leading contributions to the fidelity by different physical phenomena %
\begin{equation}
   F = %
   \left(\beta_{\gamma} \:  \beta_{\rs sc} \:  \beta_{\rs wp} \: \beta_{\rs at}\:\beta_{\rs Ry} \:\beta_{\rs tr}  \right)^2.\label{eq:FfromBeta}
\end{equation}
Here, \\
$\sbullet$
 $\beta_{\gamma}$ accounts for the spontaneous emission from the intermediate $p$-level caused by the interaction,\\
$\sbullet$  $\beta_{\rs sc}$ accounts for scattering into additional polariton channels,\\
$\sbullet$   $\beta_{\rs wp}$ includes the impact of the finite length of the wave-packets on  (a) the losses from $p$-state during propagation and (b) the inhomogeneous interaction-induced phase-shift\\
$\sbullet$     $\beta_{\rs at}$  describes the distortion of the phase shift due to spatially varying atomic density,\\
$\sbullet$ $\beta_{\rs Ry}$ denotes losses due to the finite lifetime of the Rydberg state, whereas \\
$\sbullet$ $\beta_{\rs tr}$ depicts finite beam waist effects leading to scattering of photons to other transversal photonic modes. %
}

At this point it worth mentioning the family of phase-gate schemes in  which %
at least one photon~\cite{Khazali2015,Paredes-Barato2014a,Tiarks2016,Tiarks2019,Thompson2017,Busche2017} is stored as a collective atomic excitation. %
This type of schemes is superior compared to counter-propagating scheme when it comes to the $\beta_{\rs sc} $ and $  \beta_{\rs wp} $ in equation~\eqref{eq:FfromBeta}.
However, such quantum gates suffer from an additional factor decreasing the fidelity, which accounts for the finite storage and retrieval probability of the photons;
for an efficient storage and retrieval ~\cite{Hsiao2018,Gorshkov2007,Asenjo-Garcia2017a}, such schemes can be preferable in some circumstances.

{\it Propagation inside the medium:}
First, we consider photons characterized by a single transverse mode propagating through an atomic ensemble~\cite{Bienias2016a} %
where the atomic ground state is coupled to a Rydberg $s$-state via an intermediate short-lived $p$-state, see \figref{fig1}(a).
We introduce the electric field operators $      	    \psie^{\dag}_+(z)$ and       	    $\psie^{\dag}_-(z) $ creating at position $z$ photons propagating to the right  and left, respectively.  
For the atomic density  much higher than the photonic density, the excitations of atoms 
generated by right- and   left-  moving photons into $s$-level and  $p$-level
are well-described by the bosonic field operators $	    \psis_\pm^{\dag}(z)$ and $	    \psip_\pm^{\dag}(z)$, respectively~\cite{Gorshkov2011}%
\footnote{For example, slowly varying field operators $\psip_\pm$ are defined via $\psip(z)=\psip_\pm(z) e^{\mp ik_pz+i\omega_p t}$ where $\psip$ is the field operator before going to the rotating frame whereas $\omega_p=c k_p $ is the carrier frequency of the probe field. }. 
Then, we obtain the non-interacting part of the microscopic Hamiltonian~\cite{Gorshkov2011,Bienias2014}, i.e., $H_++H_-$, under the rotating-wave approximation within the rotating frame, where
\begin{equation}
 H_{\pm} = \hbar \int \textrm{d}z
 \left(
     \begin{array}{c}
      	    \psie_\pm\\
	    \psip_\pm\\
	    \psis_\pm
     \end{array}\right)^{\dag}
     \left(\begin{array}{ccc}
   \mp i  c \partial_{z}& g    &  0 \\
         g   &   \Delta &  \Omega\\
        0  &  \Omega & 0
    \end{array} \right)
        \left(
     \begin{array}{c}
      	    \psie_\pm\\
	    \psip_\pm\\
	    \psis_\pm
     \end{array}\right).
     \label{quadraticHamiltonian-1D}
\end{equation}
Here, $g$ denotes the collective coupling of the photons to the matter via the excitation of ground state atoms into the $p$-level, 
while  $2\Omega$ denotes the Rabi frequency of the control field between the $p$-level and the Rydberg state, \figref{fig1}.
Note that the kinetic energy of the photons $\mp i \hbar \partial_{z}$ accounts for the deviation from the EIT condition.
We introduced the complex detuning  $\Delta = \delta - i \gamma$, which accounts for the detuning $\delta$ of the control field and the decay rate $2 \gamma $ from the $p$-level. 
Since we are interested in two \counterpropagating polaritons, the interaction between the Rydberg levels is described by 
\begin{equation*}
   H_{\rs rr} =  \integral{z}\!\integral{z'} V(z-z')  \psis_{\rs +}^{\dag}(z) \psis_{\rs -}^{\dag}(z') \psis_{\rs -}(z')\psis_{\rs +}(z).\nn
\end{equation*}

\begin{figure}[]
\includegraphics[width= 1\columnwidth]{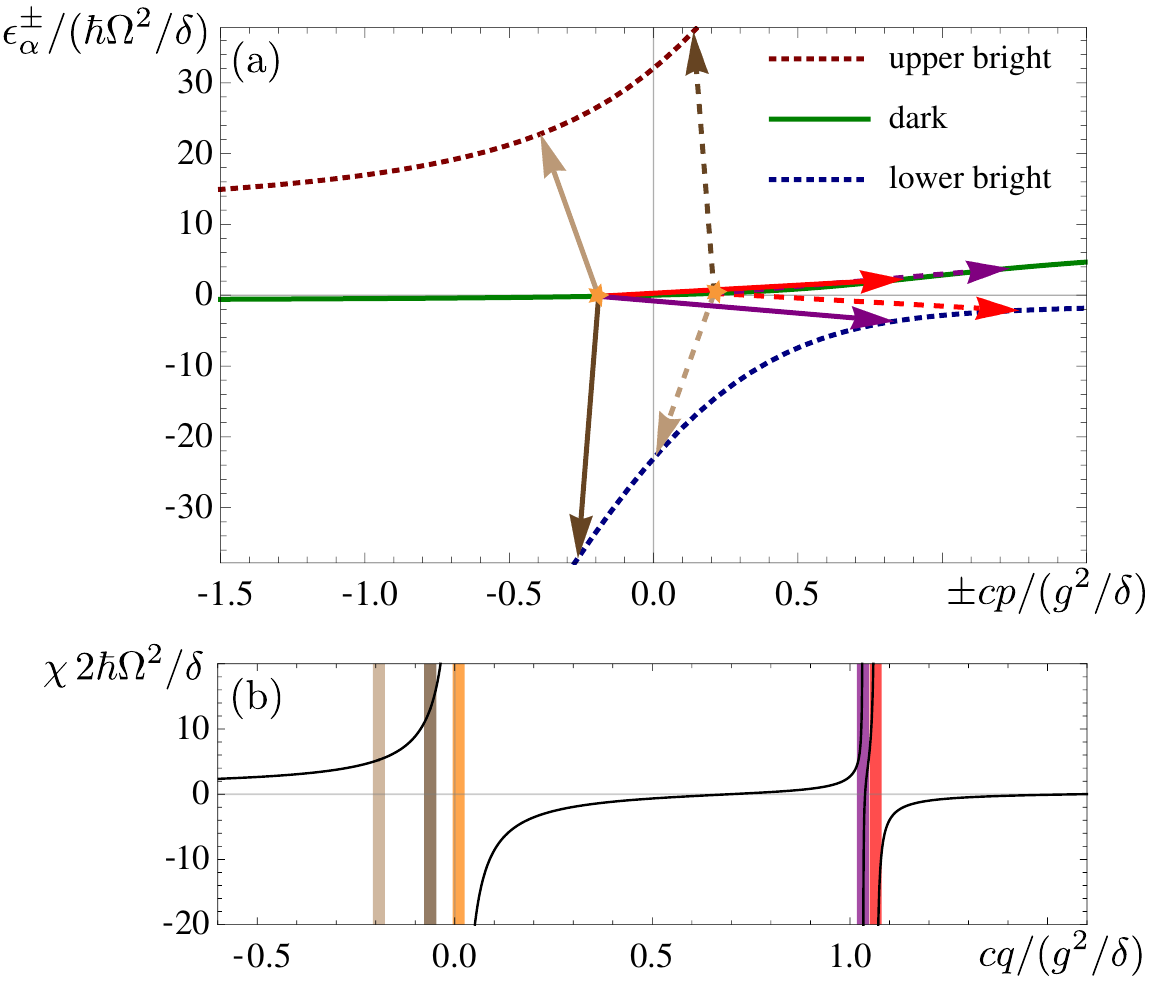}
\caption{(a) Dispersion relations for the right- and left- propagating fields are shown
 using the property that $\epsilon^{-}_\a(-p)=\epsilon^{+}_\a(p)$, see \eqref{quadraticHamiltonian-1D}, and therefore the curves overlap. 
For each direction of propagation, three noninteracting polariton branches exist. 
Two dark state polaritons (denoted by stars) can scatter into four channels, each represented by the different colour of arrows.
Solid lines correspond to right-propagating and dashed lines to left-propagating polaritons.
Note that the total momentum $\hbar K$ and energy $\hbar \omega$ are conserved during the scattering process.
Strongly suppressed losses into two bright polaritons are described by light and dark brown arrows. Crucial losses into bright-dark polariton pair are depicted by red and purple arrows. 
(b)~Full pair propagator $\chi$ in the function of relative momentum~$\hbar c q$. 
Vertical lines depict relative resonant momenta with colors corresponding to colors of arrows and stars from figure~(a). Note that both light and dark brown resonances are so narrow that cannot be resolved in the plot, and  that resonance for pair of dark polaritons (orange) is much wider than purple and red resonances.
All results are presented for $g=3\,\d,\d=3\,\Om,\gamma=0$,  $\om=0.1 \times 2\Om^2/\delta,K=-0.4\,g^2/\delta c$.
}
\label{fig2}
\end{figure}

The full scattering of the two photons inside the media
is well accounted for by the $T$-matrix.  As the interaction acts only between Rydberg states, it is sufficient to study the $T$-matrix for the Rydberg states alone, which is denoted as $T_{k k'}(K,\omega)$. Here, $\hbar K=\hbar (q_++q_-)$ denotes the center of mass momentum and $\hbar \omega$ 
the total energy, while $\hbar k'$ ($\hbar k$) is the incoming (outgoing) relative momentum. %
Note, that the total energy $\hbar \omega$ as well as the center of mass momentum $\hbar K$ are conserved in our system. The resummation of all ladder diagrams \cite{Bienias2014} leads to the integral equation
\begin{equation}
T_{kk'}(K,\omega) = V_{k-k'} + \int \frac{\textrm{d} q}{2\pi}  V_{k-q} \:  \chi_{q}(K,\omega) T_{qk'}(K,\omega).
\label{integralEq}
\end{equation}
The full pair propagator of two counter-propagating polaritons and its overlap with the Rydberg state  takes the form
\begin{equation}
\chi_q = \bar{\chi}+\frac{\alpha_{\rs D}}{\hbar k_{\rs D}(K,\omega)-\hbar q+i 0^+} + \sum_{j=1}^4\frac{\alpha_j}{\hbar k_j(K,\omega)-\hbar q+i 0^+} 
 \label{chiexpansion}.
\end{equation}
and is shown in \figref{fig2}(b).
Here, $\bar{\chi}$ accounts for the saturation of the pair propagation at large relative momenta $\hbar q\ra\pm\infty$, and takes in the relevant regime $\Omega \ll |\Delta|$ the form
\begin{equation}
\bar{\chi}(\omega) = \frac{\Delta}{2 \hbar\Omega^2} \frac{1}{1+\frac{\omega \Delta}{2 \Omega^2}}.
\end{equation}
The second term in \eqref{chiexpansion} is the pole structure for the propagation of the two incoming  dark-state polaritons.
The general expression of $\alpha_{\rs D}$ and $k_{\rs D}$ are complicated, therefore we provide below its analytical form only in the experimentally relevant regimes. The last sum accounts for the resonant scattering of the two incoming dark polaritons into four outgoing channels containing at least one 
bright  polariton, see \figref{fig2}.  
{Note that in contrast to the copropagating equation, where the massive-like behavior 
can be neglected only in certain regimes~\cite{Bienias2016}, here, the kinetic part is always  linear in the relative momentum. Thus, the large phase shift is possible without a drop of the fidelity caused by the distortion of the wave-packet shape}, as well as, we can derive analytical results.

In the following, we derive the scattering phase shift during the collision of two dark polaritons. The first important step is to notice the relation between
the scattering wave function $\psi_{k'}^{ss}(r)$ describing the amplitude to find two Rydberg excitations for the incoming relative momentum $k'$,
\begin{equation}
   \psi_{k'}^{ss}(r) = \frac{1}{V(r)} \int \frac{\textrm{d} k}{2 \pi} e^{ i k r} T_{k k'} \equiv \frac{T_{k'}(r)}{V(r)},
   \label{scatteringwavefunction}
\end{equation}
which follows from the close analogy of Eq.~(\ref{integralEq}) to the Lippmann-Schwinger equation in the scattering theory~\cite{Bienias2014}. Note that for photons of interest close to EIT resonance (i.e., $|K|,|k'|\ll g^2/|\delta |c$) and  for $|r|\gg\xi$, we have %
$\psi_{k'}^{ss}(r) = \frac{g^2}{\Omega^2}\mathcal{E}\mathcal{E}_{k'}(r) $. In the next step, we  absorb
the saturation $\bar{\chi}$ by introducing the effective interaction potential~\cite{Bienias2014}
\begin{equation}
  V^{\rs eff} (r) = \frac{V(r)}{1- \bar{\chi}(\omega) V(r)}
  \label{effectiveinteraction}.
\end{equation} 
The effective interaction potential saturates for short distances at $-1/\bar{\chi}$ and is characterized by the blockade radius $\xi$, which for a van der Waals interaction takes the from $\xi = (|C_{6} \Delta/2 \hbar \Omega^2|)^{1/6}$. Then, Eq.~(\ref{integralEq}) for the $T$-matrix simplifies to
\begin{equation}
T_{kk'} = V^{\rs eff}_{k-k'} + \int \frac{\textrm{d} q}{2\pi}  V^{\rs eff}_{k-q} \:  \big[\chi_{q} - \bar{\chi}\big]T_{qk'}
\end{equation}
and includes the contribution from all different poles in the two-particle propagator.

Finally, we focus on the experimentally 
relevant regime described by slow light polaritons  $g\gg\Om$  with large single photon detuning $|\D|\gg\Om$.
Then, the last term in Eq.~(\ref{chiexpansion}) describing scattering into a bright polaritons is strongly suppressed for small energies 
$|\omega| \leq 2\Om^2/|\D| $;  the influence of these additional processes on the gate fidelity can be discussed within perturbation theory, see below. 
In the leading order, we can therefore neglect these terms and study the $T$-matrix with the dominant pole for the propagation of the dark polaritons alone. 
Within this regime, we obtain expressions in function of $K$ and $\omega$
\begin{eqnarray*}
\alpha_{\rs D}(K,\omega) &=& \!\frac{g^2}{\Omega^2 c}\!
\left(
\frac{1}{2\sqrt{{\left(\frac{c \Delta K}{g^2}\right)^2 \!\left(\frac{\Delta  \omega }{2 \Omega ^2}+1\right)^2}\!+\!1\!}}\!+\!\frac 1 2\!\right)\! ,\\
k_{\rs D}(K,\omega) &=&\frac{ g^2}{2\Delta c}
\frac{ 1+\frac{\Delta  \omega }{\Omega ^2}- \sqrt{1+\left(\frac{c\Delta K}{g^2}\right)^2\left(\frac{\Delta  \omega }{2 \Omega ^2}+1\right)^2}}{1+  {\Delta  \omega }/{(2 \Omega ^2)}}.\nn
\end{eqnarray*}
Furthermore, we are interested in scattering processes where the incoming momentum is on-shell and describes two dark polaritons, thus, the incoming momentum $\hbar k'$ reduces to $\hbar k_{\rs D}$  with conserved energy $\hbar \omega$  and center of mass momentum $\hbar K$. 
Then, the equation for the $T$-matrix can be analytically solved and takes the form
\begin{equation}
  T_{k_{\rs D}}(r) = V^{\rs eff}(r) e^{i k_{\rs D} r} \exp\left[ - i \frac{\alpha_{\rs D}}{\hbar} \int_{-\infty}^{r} \!\!\!\!\textrm{d}z \:V^{\rs eff}(z)\right].
  \label{eq:TMatrix1D}
\end{equation}
With the relation (\ref{scatteringwavefunction}) between the scattering wave function 
and the $T$-matrix, the final result for the phase shift due to the collision of the two dark polaritons takes the form
\begin{eqnarray}
   \psi_{k_{\rs D}}^{ss}(r) &= & \frac{1}{1-\bar{\chi} V(r)}e^{i k_{\rs D} r} \exp\left[ - i \frac{\alpha_{\rs D}}{\hbar} \int_{-\infty}^{r} \textrm{d}z V^{\rs eff}(z)\right]  \nonumber \\
   & = & \left\{ 
   \begin{array}{ccc}
     e^{i k_{\rs D} r}  &  r \rightarrow & - \infty \\
     e^{i \varphi} e^{i k_{\rs D} r}   & r \rightarrow & \infty
   \end{array}\right.
   \label{analyticalSol}
\end{eqnarray}
with the phase shift
\begin{equation}
\varphi = - \frac{\alpha_{\rs D}}{\hbar}\int_{-\infty}^{\infty} \textrm{d}z\, V^{\rs eff}(z).
\end{equation}
Note, that the wave function is strongly suppressed within the blockade region $|r| < \xi$ due to the blockade phenomena quenching two Rydberg excitations at short distances. In the presented leading order, the scattering into additional channels is suppressed and therefore we obtain only a phase shift. The phase shift contains an imaginary part accounting for the losses during the scattering process by spontaneous emission from the intermediate $p$-level.
Performing the remaining integration for a microscopic van der Waals interaction, the phase reduces to
\begin{equation} \vspace{-0em}
\varphi(K,\omega)=\alpha_{\rs  D}(K,\omega)\frac{2  \pi  g^2 \xi (- \textrm{sgn} [{C_6}] \Delta)^{1/6}  }{3 c {\left| \Delta \right| }^{1/6} {\Delta  \left(\frac{\Delta  \omega }{2\Omega ^2}+1\right)^{7/6}} }\,.
\label{eq:phaseFactorMono}
\vspace{-0.em} \end{equation}
Note, that the phase still depends on the center of mass momentum and the total energy, and therefore, we expect a weakly 
inhomogeneous phase shift for a wave packet. The leading contribution is determined  on the two-photon resonance, i.e., $\om=0$ and $K=0$, where the solution agrees with \cite{Gorshkov2011b}. 
For $\gamma\ll|\delta|$ the exponent $\varphi$ simplifies to %
\begin{equation}
\phi+i\eta=\frac{2  \pi  g^2 \xi}{3\delta c}  \left(1+i\frac 5 6 \frac \gamma \d\right)
	\label{eq:phaseFactorMono0},
\end{equation}
where  $\phi$ describes the coherent phase shift and can be expressed using optical depth $\kappa_\xi$ per blockade 
radius, $\phi=(\pi/3)\kappa_\xi\gamma/\delta$. 
{In turn $\eta$ denotes the losses via spontaneous emission from the $p$-level
and therefore provides a reduction of the outgoing wave function, i.e.,
\begin{equation}
\beta_{\gamma}^2= e^{- 2 \eta} \approx 1- 2 \eta\,.
\end{equation}
The $\beta_{\gamma}$ is the main contribution to the reduction of the gate fidelity, and it requires the
far detuned regime $|\delta|\gg \gamma$ in combination with large optical depth per blockade radius. }

{\it Scattering into additional open channels---}
Next, we continue with a detailed analysis of all additional contributions leading to a reduction of the gate fidelity. 
We start with the study of the influence of the additional poles in the two polariton propagator $\chi$, see \figref{fig2}. These poles describe additional open
channels characterized by the relative momenta $k_j$. Therefore, the interaction between the polaritons leads to  scattering of the
incoming dark polaritons into these open channels. 
Consequently, the outgoing wave function takes the form %
\begin{equation}
   \psi^{ss}_{k_{\rs D}}(r) =  \beta_{\rs sc} e^{i \varphi} e^{i k_{\rs D} r} + \sum_{j=1}^{4} \epsilon_{j} e^{i k_j r},
\end{equation}
with $|\beta_{\rs sc}|^2$ a reduced  probability to remain in the dark polariton state, and $|\epsilon_j|^2$ a probability to be in the other channel. 
The analysis can be performed straightforwardly in lowest order perturbation theory in $\alpha_{j}/\alpha_{\rs D}$.
Then, the correction $\delta T$ to the $T$-matrix is determined by the equation
\begin{eqnarray}
\delta T_{k_{\rs D}}(r)&=& - \frac{i \alpha_{\rs D}}{\hbar} V^{\rs eff}(r) e^{i k_{\rs D} r} \int_{-\infty}^{r} \!\!\!\!\! \textrm{d}z e^{- ik_{\rs D} z} \delta T_{k_{\rs D}}(z) \\
& &- \sum_{j=1}^{4} \frac{ i \alpha_{j}}{\hbar}   V^{\rs eff}(r)\int_{-\infty}^{r} \textrm{d}z e^{i k_{j}(r-z)} T_{k_{\rs D}} (z) \nonumber.
\end{eqnarray}
The general solution takes the form %
\begin{eqnarray}
 \delta T_{k_{\rs D}}(r)  =  \sum_{j=1}^{4} \frac{ \alpha_{j}}{\alpha_{\rs D}} \left\{  e^{i k_{j} r} V^{\rs eff}(r) \int_{-\infty}^{r} \!\!\!\!\! \textrm{d}z e^{i (k_{\rs D} -k_{j}) z}  \partial_{z} e^{i \theta(z)}  \right. \nonumber\\
 \hspace{0pt}- \left. T_{k_{\rs D}}(r) \int_{-\infty}^{r} \!\!\!\!\! \textrm{d}z  \int_{-\infty}^{z} \!\!\!\!\! \textrm{d}y e^{i (k_{j}-k_{\rs D})(z-y)} \partial_{z} e^{- i \theta(z)}  \partial_{y} e^{i \theta(y)}     \right\} \nonumber,
\end{eqnarray} 
where %
 $\theta(z) =- \alpha_{\rs D}\int_{-\infty}^{z} dy V^{\rs eff}(y)/\hbar $.
The first term describes the weight in the additional open channels, i.e.,
\begin{equation}
\epsilon_{j} = \frac{\alpha_{j}}{\alpha_{\rs D}} \int_{-\infty}^{\infty} \!\!\!\!\! \textrm{d}z \: \:   e^{i (k_{\rs D} - k_{j}) z} \partial_{z} e^{i \theta(z)}
\end{equation}
whereas the second term accounts for the reduction of the outgoing dark polariton state, and can be written in leading order 
 $\alpha_{i}/\alpha_{\rs D}$ as
\begin{eqnarray*}
1-|\beta_{\rs sc}|^2=\hspace{19em}\\ 2\sum_{j=1}^{4} \frac{ \alpha_{j}}{\alpha_{\rs D}}  \left|
\int_{-\infty}^{r} \!\!\!\!\! \textrm{d}z  \int_{-\infty}^{z} \!\!\!\!\! \textrm{d}y e^{i (k_{j}-k_{\rs D})(z-y)} \partial_{z} e^{- i \theta(z)}  \partial_{y} e^{i \theta(y)} 
      \right|.  \nn
\end{eqnarray*}

In order to estimate the value of the suppression, it is sufficient to focus on the resonant regime with $\omega =0$ and $K=0$. Then, the values of momenta $k_{j}$ and the weight of the residues $\alpha_j$ take a simple analytical form.  Specifically, in the experimentally relevant regime with $ \Omega \ll g$, we find that the poles group into pairs of equal weight
\begin{eqnarray}
   k_{1} =k_{2} = \frac{g^2}{\Delta c} , &\hspace{20 pt}  & \alpha_{1} = \alpha_{2} = \frac{g^2+\Delta^2}{4 \Delta^2 c},\\
    k_{3} =k_{4} = -\frac{\Delta}{c} ,&\hspace{20 pt}  & \alpha_{3} = \alpha_{4} = \frac{g^2 \Omega^4}{4 (g^2+\Delta^2)^3 c}.
\end{eqnarray}
The first two poles describe  excitations at high relative momentum, whereas the second pair of poles at low relative momenta which is always strongly suppressed for slow light polaritons, see \figref{fig2}.
Therefore, the reduction due to the scattering into the additional open channels can be estimated to satisfy the inequality
\begin{equation}
\label{corrEstim}
|\beta_{\rs sc}|^2 \gtrsim 1-\phi^2 |2 \alpha_{1}/\alpha_{\rs D}|= 1-  \phi^2\frac{\Omega^2}{|\Delta|^2}\frac{|\Delta|^2+g^2}{2g^2} .
\end{equation}
This term is suppressed for $\Omega \ll |\Delta|$, which provides an additional constraint on
the experimental parameters.

{\it Wave packet propagation---} The next step is to study the influence of realistic wave packets onto the gate fidelity. There are two different contributions: First one is a distortion of the wave packet due to the non-linear corrections to the  dispersion relation of the dark polariton, which is also present for noninteracting polaritons. 
Second contributions comes from  the  dependence of phase factor $\varphi$ on $ K$ and $\omega$ %
leading to the longitudinally inhomogeneous phase shift for wave packets. %
In the following, both phenomena will be discussed in detail.

The right and left moving photonic wave packets inside the homogeneous atomic media are described by the slowly varying envelope for the
electric field, 
\begin{equation}
\psieWf_{\rs \pm}(z,t)=
\int \frac{\textrm{d} \nu}{2 \pi}
\psieWf_\pm(\nu)
\exp[ \pm i p(\nu)  (z - L_{\rs \pm})- i \nu t] .
\label{photonShape}
\end{equation}
Here, $\psieWf_{\rs \pm}(L_{\rs \pm},t)$ describes the electric field of the incoming photonic wave packet at the boundary of the atomic media with $L= L_{\rs +}\!-\!L_{\rs -}$
the length of the media, see \figref{fig1}. %
In addition, the momentum $\hbar p(\nu)$ of a polariton is related to frequency $\nu$  by the dispersion relation for dark polaritons. 
The momentum exhibits the general form
\begin{equation}
   p(\nu) = \nu  \: \frac{g^2 + \Omega^2 + \Delta \nu - \nu^2}{c \left( \Omega^2 + \Delta \nu - \nu^2\right)}\approx \frac{\nu}{v_{g}} -\frac{\hbar \nu^2}{2 m v_g^3}, \label{momentum}
\end{equation}
which in the experimentally relevant regime close to the EIT resonance is well described  by the slow light velocity $v_{g} = c \Omega^2/g^2$ and the effective mass $m = \hbar g^4/(2 c^2 \Delta \Omega^2)$. 
The mass leads to a wave packet dispersion as well as losses  from the intermediate $p$-level due to the imaginary part of the mass. 
Note that even though it looks as if the mass term  contributes to the dynamics of  the pair of polaritons, the physics is more subtle. 
Namely, in \eqref{chiexpansion} we showed that two-photon-dynamics for conserved $K,\omega$ has a linear dependence  %
on the relative momentum
[which allows simple analytical solution, \eqref{eq:TMatrix1D}]. However, a conserved relative momentum $k_D$ is a function of $K,\omega$ and $m$ 
which corresponds to~\eqref{momentum} which we analyze here.

The probability for the polariton to pass through the media is  given by
\begin{equation}
    P_{\pm} = \int \frac{d \nu}{2 \pi}   |\psieWf_{\rs \pm}(\nu)|^2  e^{- 2 p''(\nu) L}\,,
\end{equation}
where $p''(\nu)$ denotes the imaginary part of $p(\nu)$. %
Therefore, the dipolar interaction between two polaritons leads to two effects:  first, a weak shift for the scattering phase $\Delta \phi$ 
due to the  averaging over different energy and momentum states, secondly, a reduction in the gate fidelity
$\beta_{\rs wp}$. Both quantities are determined by the equation
\begin{eqnarray}
\beta_{\rs wp}e^{i \Delta \phi} &= &
\int \frac{\textrm{d} \nu_1 \textrm{d} \nu_2}{(2\pi)^2 }
|\psieWf_{\rs +}(\nu_1)\psieWf_{\rs -}(\nu_2)|^2 
e^{- 2L [ p''(\nu_{1}) + p''(\nu_{2})] } \nonumber \\ & &
\times \exp\left[i \varphi(K,\omega)  - i \phi + \eta)\right]  .
\end{eqnarray}
It is important that $\varphi(K,\omega)$ depends on the total energy $\hbar\omega= \hbar\nu_1+\hbar\nu_2$ 
and the center of mass momentum $\hbar K=\hbar p(\nu_{1}) - \hbar p(\nu_{2})$.

In order to illustrate the phenomena of the wave packet propagation, we focus on a Gaussian incoming wave packet
\begin{equation}
\psieWf_\pm(\nu) =\frac{\exp \left[ - \nu^2/2 \sigma^2\right]}{(\pi \sigma^2)^{1/4}}
\end{equation}
with a small bandwith %
$\sigma  \ll \Omega^2/|\Delta|$. Then, it is sufficient to keep the
leading quadratic corrections in $\omega$ and $K$. 
First, the impact on the fidelity simplifies to %
\begin{equation}
\beta_{\rs wp}=  P_{\rs +} P_{\rs -}  \exp \left[ -
	\frac{49}{2}\phi^2  \left( \frac{\sigma |\Delta|}{12 \Omega^2} \right)^2  \right],
	\label{eq:betaWP}
\end{equation}
with
\begin{equation}
P_{\rs \pm} =  \exp \left[ -2 \frac{Lg^2}{\delta c} \left(\frac{\delta \sigma}{\Om^2}\right)^2\frac{\gamma}{\delta}
 \right].
\end{equation}
Second, the shift in the coherent phase reduces to 
\begin{equation}
\Delta \phi=   \frac{19}{2}  \phi \left(   \frac{\sigma |\Delta|}{12\Omega^2}\right)^2 .
\end{equation}

{\it Impact of atomic distribution}---
Additional contribution to the fidelity comes from the inhomogeneity of the atomic distribution. To illustrate this effect, we consider Gaussian atomic distribution $n_{\rs at}(z) =n_0\exp \left[ - z^2/ \mathcal{L}^2\right]$, where $\mathcal{L}$ characterizes the cloud's length. 
 In the relevant regime of $\xi\ll \mathcal{L}$,  %
the scattering phase shift depends locally on the atomic density at which two polaritons interact with each other. 
Inside the medium photons are compressed to the size $\s_z=c\Omega^2/(g^2\sigma)$, where $g$ in this paragraph denotes the value of collective coupling at the center of the cloud. 
For short photons $\s_z \ll\mathcal{L}$%
, the corrections to the fidelity and phase shift take the form~\cite{SuppBienias} 
\begin{equation}
\beta_{\rs at}e^{i \Delta \phi_{\rs at}} =1 -i\phi \left(\frac{    \sigma_z}{2 \mathcal{L}}\right)^2 -\frac{3}{2}\phi ^2\left(\frac{ \sigma_z}{2 \mathcal{L}}\right)^4 \,.\label{longitudinal}
\end{equation}
This puts additional constraint on the length of the medium: $\mathcal{L}\gg\s_z $.

\textit{Transversal size of the photons}---
\begin{figure}[t]
\includegraphics[width= \size\columnwidth]{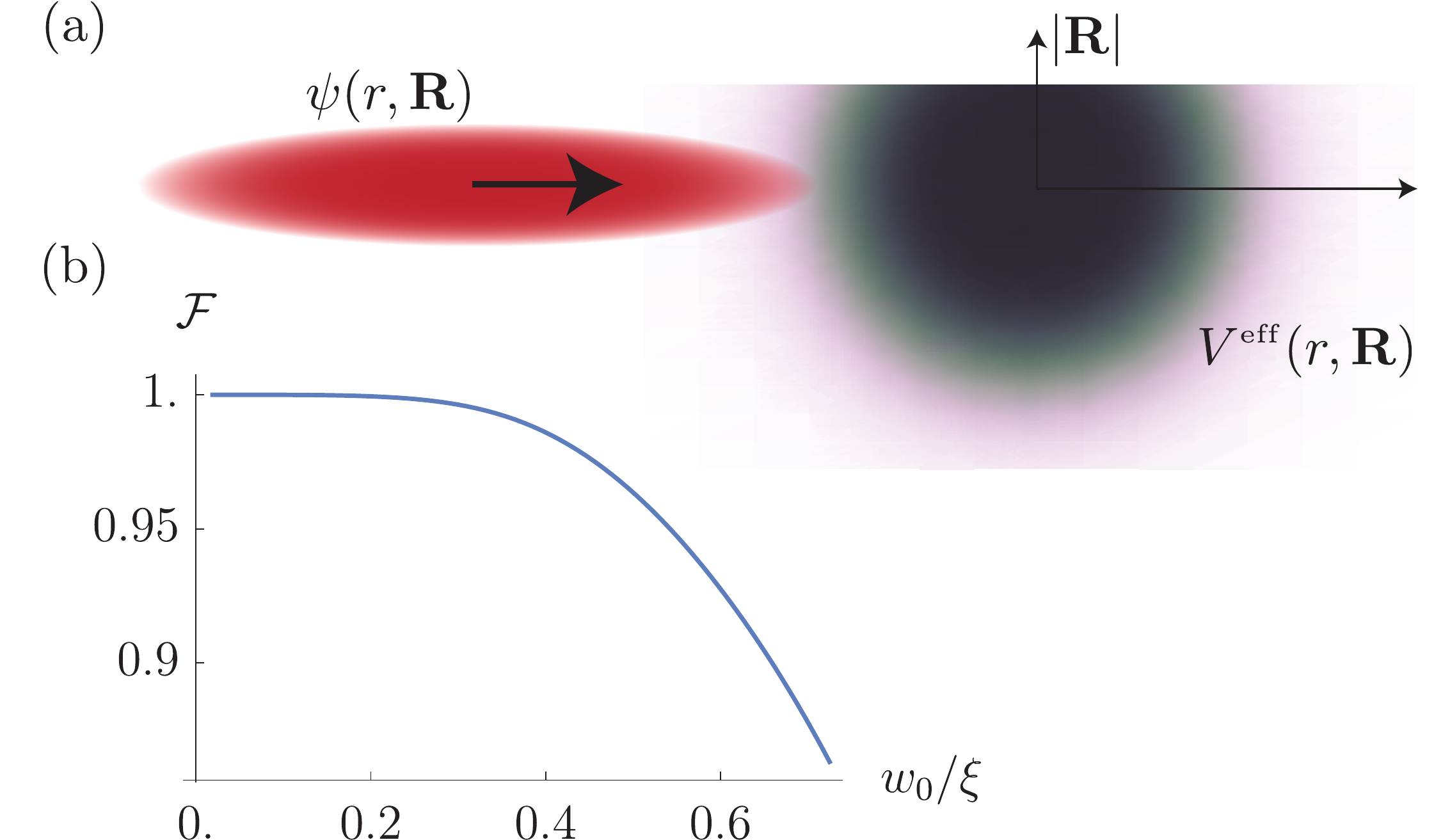}
\caption{
(a) Two-photon wave-function,  having finite transverse size, scatters in center-of-mass (i.e., $z_-+z_+=\textrm{const}.$ and $\bR_++\bR_-=\textrm{const}.$) frame on the effective potential $\Veff(r,\bR)$, where $\bR=\bR_+-\bR_-$. 
(b)~Fidelity as a function of a transverse width $\s_{\perp}$ for the phase shift $\phi=\pi/2$: The drop of the fidelity for $\sRp/\xi<0.35$ is negligible. Intuitively, due to the plateauing of $\Veff$, photons experience a homogeneous transversal phase shift. 
}
\label{fig0}
\end{figure}
Two colliding wave-packets in the lowest Laguerre-Gauss mode %
are described by $\psieWf_\pm(z_\pm,\bR_\pm)=\psieWf(z_\pm)u(\bR_\pm)$ with  $u(\bR)=\exp[-{(x^2+y^2)}/{\sRp^2}]\sqrt{2/\pi}/ \sRp$ where $\sRp$ is the probe beam waist.
The interaction potential generalized to a quasi-1D geometry  depends on the relative transversal distance, $\Veff(r,\bR)=V(\br)/(1- \bar{\chi}(\omega) V(\br))$, \cite{SuppBienias}. %
For beam waist comparable to the blockade radius, this leads to the transversely inhomogeneous phase shift 
 \cite{Friedler2005,He2011,Moos2015}. 
To estimate the correction to the fidelity, we neglect transversal part of the photonic paraxial wave equation~\cite{SuppBienias}, which leads to $\beta_{\rs tr}$ being equal to
\begin{equation*} 
\vspace{-0em}
\Big{|}
\integral{\bR_{\rs+}}\hspace{-0.2em}\textrm{d}{\bR_{\rs-}}
|u(\bR_{\rs+})u(\bR_{\rs-})|^2\exp [i\varphi (|\bR_{\rs+}-\bR_{\rs-}|,\xi)]\Big{|}\,.
\vspace{-0.em}
\end{equation*}
From $\beta_{tr}$ in the limit of $\gamma/\delta\ra 0$ [\figref{fig0}(b)] we see that
the drop of fidelity can be neglected for $\sRp/\xi\lesssim 0.35$, which for exemplary experimental parameters, $\sRp=4.2\,\mu $m, $|nS\rangle=|100S\rangle$, $\Om=\gamma$, and $\d=5\gamma$, gives $\sRp/\xi=0.17$. %
Intuitively, the reason for this behavior is that polaritons interact via $\Veff(\br) $ %
which is nearly constant for the distances shorter than the blockade radius. 
It is an important feature of Rydberg polaritons resulting in the phase shift being nearly homogeneous also in the transverse direction. 
Note that this feature is no longer present in the proposals based on the pulses propagating in two parallel but spatially-separated  photonic modes~\cite{He2014a} and should be carefully taken into account.

{\it Rydberg lifetime ---}can be estimated using delay time $L/v_g$ the polariton spends inside the medium during which it is primarily of Rydberg character, which leads to   $\beta_{\rs Ry}=\exp[ -4\gamma_s \frac{L}{v_g}]$.

{\it Experimental realization}---
\begin{figure}[t]
\center
\includegraphics[width= 0.31\textwidth]{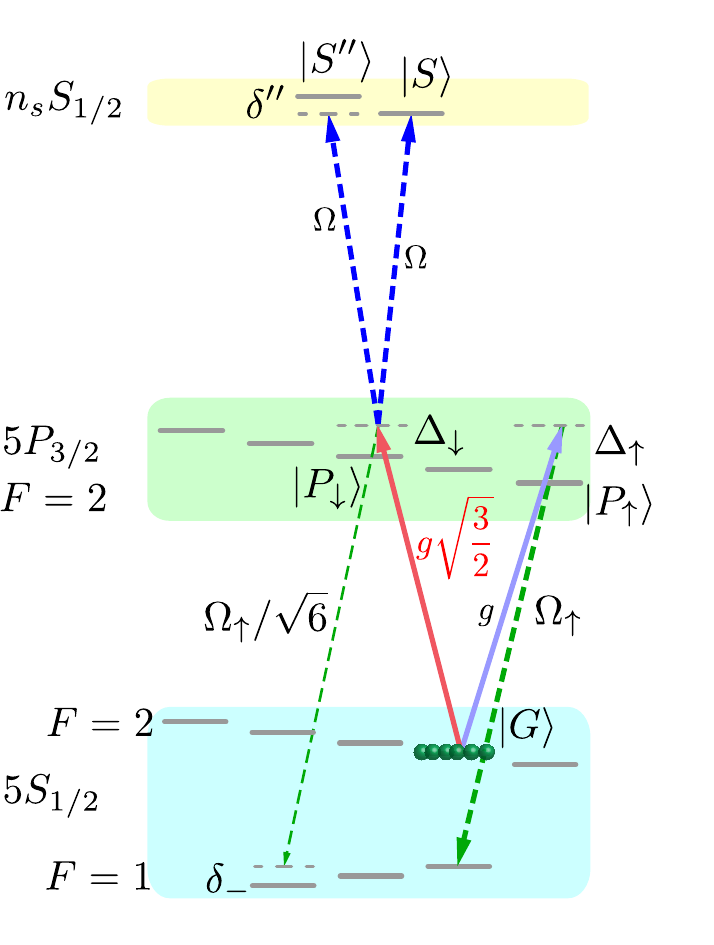}
\caption{
Atomic scheme.
Assuming repulsive vdW interactions and choosing $\delta''\gg\Omega^2/|\Delta_{\uparrow}|$ as well as $\delta_\downarrow<0$ (where $\Delta_\downarrow=\delta_\down-i\gamma$), we avoid Raman resonance for $\ket{S}$ and $\ket{S''}$, as well as, can neglect coupling to $\ket{S''}$ at all. See main text for additional description.
}
\label{fig:atomicScheme}
\end{figure}
Two-qubit conditional quantum phase gate (QPG)~\cite{Lloyd1995,Turchette1995,Rauschenbeutel1999} is universal because combined with rotations of individual qubits enables any quantum computation.
The QPG transformation reads
\beqa
\ket{b,j} \ra \exp[i\phi\, \delta_{b,\down}\delta_{j,\down}] \ket{b,j} \label{eq:QPGdefinition},
\eeqa
where $\ket{b},\ket{j}$ depict basis states $\ket{\up}$ and $\ket{\down}$ of the two qubits of interest and $\delta_{b,\down}\delta_{j,\down}$ are standard Kronecker symbols.
It is important to stress that QPG leaves basis states unchanged except a homogeneous phase shift $\phi$ when both are in $\ket{\down}$ state.
In order to implement the gate, it is important that both states $\ket{\up}$ and $\ket{\down}$ acquire the same time delay inside the medium, but only the state $\ket{\down}$
becomes a Rydberg polariton.  In \figref{fig:atomicScheme}, we show the atomic scheme which satisfies this condition. %
As a ground state we take
$|g\rangle=|5^2S_{1/2},F\!=\!2,m_F\!=\!1\rangle$
from which we couple with $\sigma_+$ polarized photons to 
$|p_\up\rangle=|5^2P_{3/2},F\!=\!2,m_F\!=\!2\rangle$, 
and with $\sigma_-$-polarized to
$|p_\down\rangle=|5^2P_{3/2},F\!=\!2,m_F\!=\!0\rangle$.
We choose control field to be  $\sigma_+$ polarized, which ensures that coupling to any Rydberg $s$-states  from $\ket{p_\up}$ is zero because 
$
|p_\up\rangle=\frac{1}{\sqrt{2}}\left[|{\textstyle m_j=\frac{1}{2}}\rangle|m_I={\textstyle\frac{3}{2}}\rangle-|m_j={\frac{3}{2}}\rangle|m_I=-{\textstyle\frac{1}{2}}\rangle\right].\nn$
Whereas  from $|p_\down\rangle$ we can couple
 with $\sigma_+$ control field to two different $s$-states (because
$%
|p_\down\rangle=%
\frac{1}{2}\left[|{\textstyle m_j=-\frac{3}{2}}\rangle|m_I={\textstyle\frac{3}{2}}\rangle
+|{\textstyle m_j=-\frac{1}{2}}\rangle|m_I={\textstyle\frac{1}{2}}\rangle\right.$\\ %
$-\left. |{\textstyle m_j=\frac{1}{2}}\rangle|m_I=-{\textstyle\frac{1}{2}}\rangle
-|{\textstyle m_j=\frac{3}{2}}\rangle|m_I=-{\textstyle\frac{3}{2}}\rangle
\right]\nn
$%
), from which we single out one  (i.e., $\ket{S}$ in Fig.~\ref{fig:atomicScheme}) by applying external magnetic field.
For $|\d_-\Delta_\down|\gg \Om^2$ and $\Om_\up=\Om\sqrt{2/3}$ both polaritons have the same group velocity, therefore, the time delay inside the medium is the same. 
This way the only impact of the interaction on the qubit state is the phase shift for $\ket{\downarrow,\downarrow}$ state, which corresponds to $\ket{\psie_+\psie_-}$ photons in our analysis of the phase gate fidelity.

\begin{figure}[h]
\includegraphics[width= 0.85\columnwidth]{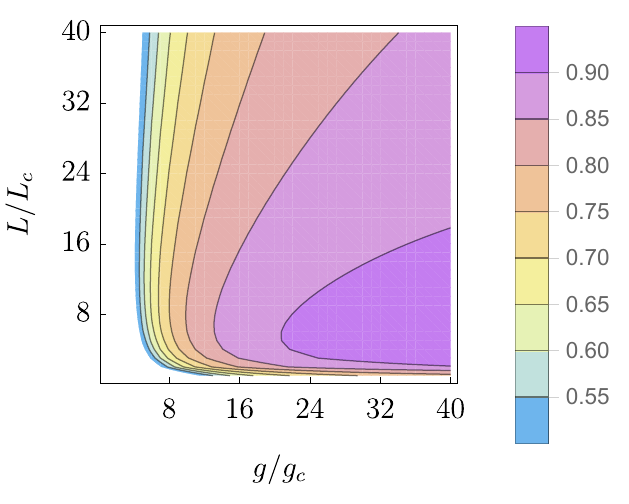}
\caption{
The $\sqrt{F}$ including all presented corrections for $\phi=\pi/2$ as a function of collective coupling $g$ and length of the medium $L$, with parameters characteristic for experiments presented in Refs~\cite{Firstenberg2013,Gorniaczyk2014}:
$L_c=160\mu m$ and $g_c/2\pi =4.4 $GHz. %
All our estimates assume that the corrections are small and a pulse is long, $\s_z g^2/|\D|c\gg 1$. Therefore, the estimates are not valid in the regimes where $F$ drops much below 1, therefore, we do not plot fidelity there. 
}
\label{fig:estimates}
\end{figure}

{\it Estimate of the optimal parameters for the phase gate}---Based on the estimates of all the detrimental effects, we numerically find optimal parameters $\d$ and $\Om$ for fixed $C_6$, $g$, %
$L$ and %
 $\phi$ in case of $^{87}$Rb atoms (for which $\gamma/\pi = 6.1\,$MHz). A resulting fidelity for $\ket{100S}$ state (for which $\gamma_s/2\pi= 2.3$\,kHz) and $\phi=\pi/2$ as a function of  $g$ and $L=2\mathcal{L}$ is presented in \figref{fig:estimates}.  
In order to neglect interaction effects for the entering and leaving the medium, as well as to ensure that photons will completely interact with each other, we set $\s_z=L/8$.
Some of the corrections are overestimated, like scattering to other channels, see \eqref{corrEstim}, and thus true fidelity can be higher. 
Presented \figref{fig:estimates} shows that the fidelity of a phase gate is not constrained fundamentally. %
Note that our analysis is a good starting point to find optimal parameters using full two-photon propagation numerics.

{\it Acknowledgments}---
We thank S. Hofferberth and A.V. Gorshkov  for discussions. 
We acknowledge support by the European Union under the ERC consolidator grant SIRPOL (grant N. 681208).
P.B. acknowledges support by 
 by ARL CDQI, ARO MURI, AFOSR, NSF PFC at JQI, DoE ASCR Quantum Testbed Pathfinder Program (award No. DE-SC0019040), DoE BES QIS program (award No. DE-SC0019449), and NSF PFCQC program.
\bibliography{library}

\begin{thebibliography}{10}

\bibitem{Nielsen2000}
M.~A. Nielsen and I.~L. Chuang, \emph{{Quantum computation and quantum
  information}} (2000).

\bibitem{Turchette1995}
Q.~A. Turchette, C.~J. Hood, W.~Lange, H.~Mabuchi and H.~J. Kimble,
  \emph{{Measurement of conditional phase shifts for quantum logic}}, Phys.
  Rev. Lett., \textbf{75}, 4710--4713 (1995).

\bibitem{Fushman2008}
I.~Fushman, D.~Englund, A.~Faraon, N.~Stoltz, P.~Petroff and J.~Vuckovic,
  \emph{{Controlled phase shifts with a single quantum dot.}}, Science,
  \textbf{320}, 769--72 (2008).

\bibitem{Parigi2012}
V.~Parigi, E.~Bimbard, J.~Stanojevic, A.~J. Hilliard, F.~Nogrette,
  R.~Tualle-Brouri, A.~Ourjoumtsev and P.~Grangier, \emph{{Observation and
  Measurement of Interaction-Induced Dispersive Optical Nonlinearities in an
  Ensemble of Cold Rydberg Atoms}}, Phys. Rev. Lett., \textbf{109}, 233602
  (2012).

\bibitem{Volz2014}
J.~Volz, M.~Scheucher, C.~Junge and A.~Rauschenbeutel, \emph{{Nonlinear $\pi$
  phase shift for single fibre-guided photons interacting with a single
  resonator-enhanced atom}}, Nat Phot., \textbf{8}, 965--970 (2014).

\bibitem{Reiserer2014}
A.~Reiserer, N.~Kalb, G.~Rempe and S.~Ritter, \emph{{A quantum gate between a
  flying optical photon and a single trapped atom.}}, Nature, \textbf{508},
  237--40 (2014).

\bibitem{Fleischhauer2005}
M.~Fleischhauer, \emph{{Electromagnetically induced transparency: Optics in
  coherent media}}, Rev. Mod. Phys., \textbf{77}, 633--673 (2005).

\bibitem{Lo2011}
H.-Y. Lo, Y.-C. Chen, P.-C. Su, H.-C. Chen, J.-X. Chen, Y.-C. Chen, I.~a. Yu
  and Y.-F. Chen, \emph{{Electromagnetically-induced-transparency-based
  cross-phase-modulation at attojoule levels}}, Phys. Rev. A, \textbf{83},
  041804 (2011).

\bibitem{Shiau2011}
B.~W. Shiau, M.~C. Wu, C.~C. Lin and Y.~C. Chen, \emph{{Low-light-level
  cross-phase modulation with double slow light pulses}}, Phys. Rev. Lett.,
  \textbf{106}, 193006 (2011).

\bibitem{Feizpour2015}
A.~Feizpour, M.~Hallaji, G.~Dmochowski and A.~M. Steinberg, \emph{{Observation
  of the nonlinear phase shift due to single post-selected photons}}, Nat.
  Phys., \textbf{11}, 905--909 (2015).

\bibitem{Nemoto2004}
K.~Nemoto and W.~J. Munro, \emph{{Nearly deterministic linear optical
  controlled-NOT gate}}, Phys. Rev. Lett., \textbf{93}, 250502 (2004).

\bibitem{Mohapatra2007}
A.~K. Mohapatra, T.~R. Jackson and C.~S. Adams, \emph{{Coherent Optical
  Detection of Highly Excited Rydberg States Using Electromagnetically Induced
  Transparency}}, Phys. Rev. Lett., \textbf{98}, 113003 (2007).

\bibitem{Pritchard2010}
J.~D. Pritchard, D.~Maxwell, A.~Gauguet, K.~Weatherill, M.~Jones and C.~Adams,
  \emph{{Cooperative Atom-Light Interaction in a Blockaded Rydberg Ensemble}},
  Phys. Rev. Lett., \textbf{105}, 193603 (2010).

\bibitem{Mohapatra2008}
A.~K. Mohapatra, M.~G. Bason, B.~Butscher, K.~J. Weatherill and C.~S. Adams,
  \emph{{A giant electro-optic effect using polarizable dark states}}, Nat.
  Phys., \textbf{4}, 890--894 (2008).

\bibitem{Peyronel2012}
T.~Peyronel, O.~Firstenberg, Q.-Y. Liang, S.~Hofferberth, A.~V. Gorshkov,
  T.~Pohl, M.~D. Lukin and V.~Vuleti{\'{c}}, \emph{{Quantum nonlinear optics
  with single photons enabled by strongly interacting atoms.}}, Nature,
  \textbf{488}, 57--60 (2012).

\bibitem{Dudin2012}
Y.~O. Dudin and A.~Kuzmich, \emph{{Strongly interacting Rydberg excitations of
  a cold atomic gas.}}, Science, \textbf{336}, 887--9 (2012).

\bibitem{Dudin2012b}
Y.~O. Dudin, F.~Bariani and a.~Kuzmich, \emph{{Emergence of Spatial Spin-Wave
  Correlations in a Cold Atomic Gas}}, Phys. Rev. Lett., \textbf{109}, 133602
  (2012).

\bibitem{Maxwell2013}
D.~Maxwell, D.~J. Szwer, D.~P. Barato, H.~Busche, J.~D. Pritchard, A.~Gauguet,
  K.~J. Weatherill, M.~P.~A. Jones and C.~S. Adams, \emph{{Storage and Control
  of Optical Photons Using Rydberg Polaritons}}, Phys. Rev. Lett.,
  \textbf{110}, 103001 (2013).

\bibitem{Hofmann2013}
C.~S. Hofmann, G.~G{\"{u}}nter, H.~Schempp, M.~Robert-de Saint-Vincent,
  M.~G{\"{a}}rttner, J.~Evers, S.~Whitlock and M.~Weidem{\"{u}}ller,
  \emph{{Sub-Poissonian Statistics of Rydberg-Interacting Dark-State
  Polaritons}}, Phys. Rev. Lett., \textbf{110}, 203601 (2013).

\bibitem{Firstenberg2013}
O.~Firstenberg, T.~Peyronel, Q.-Y. Liang, A.~V. Gorshkov, M.~D. Lukin and
  V.~Vuleti{\'{c}}, \emph{{Attractive photons in a quantum nonlinear medium.}},
  Nature, \textbf{502}, 71--75 (2013).

\bibitem{Gorniaczyk2014}
H.~Gorniaczyk, C.~Tresp, J.~Schmidt, H.~Fedder and S.~Hofferberth,
  \emph{{Single Photon Transistor Mediated by Inter-State Rydberg
  Interaction}}, Phys. Rev. Lett., \textbf{113}, 053601 (2014).

\bibitem{Baur2014}
S.~Baur, D.~Tiarks, G.~Rempe and S.~D{\"{u}}rr, \emph{{Single-Photon Switch
  based on Rydberg Blockade}}, Phys. Lev. Lett., \textbf{112}, 073901 (2014).

\bibitem{Tiarks2014}
D.~Tiarks, S.~Baur, K.~Schneider, S.~D{\"{u}}rr and G.~Rempe,
  \emph{{Single-Photon Transistor Using a F{{\"{o}}}rster Resonance}}, Phys.
  Rev. Lett., \textbf{113}, 053602 (2014).

\bibitem{Tresp2015}
C.~Tresp, P.~Bienias, S.~Weber, H.~Gorniaczyk, I.~Mirgorodskiy, H.~P.
  B{\"{u}}chler and S.~Hofferberth, \emph{{Dipolar Dephasing of Rydberg $D$
  -State Polaritons}}, Phys. Rev. Lett., \textbf{115}, 083602 (2015).

\bibitem{Gorniaczyk2016}
H.~Gorniaczyk, C.~Tresp, P.~Bienias, A.~Paris-Mandoki, W.~Li, I.~Mirgorodskiy,
  H.~P. B{\"{u}}chler, I.~Lesanovsky and S.~Hofferberth, \emph{{Enhancement of
  Rydberg-mediated single-photon nonlinearities by electrically tuned
  F{\"{o}}rster resonances}}, Nat. Commun., \textbf{7}, 12480 (2016).

\bibitem{Tresp2016}
C.~Tresp, C.~Zimmer, I.~Mirgorodskiy, H.~Gorniaczyk, A.~Paris-Mandoki and
  S.~Hofferberth, \emph{{Single-photon absorber based on strongly interacting
  Rydberg atoms}}, PRL, \textbf{117}, 223001 (2016).

\bibitem{Tiarks2016}
D.~Tiarks, S.~Schmidt, G.~Rempe and S.~Durr, \emph{{Optical phase shift created
  with a single-photon pulse}}, Sci. Adv., \textbf{2}, e1600036--e1600036
  (2016).

\bibitem{Schine2016}
N.~Schine, A.~Ryou, A.~Gromov, A.~Sommer, J.~Simon, E.~Hamiltonian and
  L.~Gauss, \emph{{Synthetic Landau levels for photons}}, Nature, \textbf{534},
  7609 (2016).

\bibitem{Sevincli2011}
S.~Sevin{\c{c}}li, N.~Henkel, C.~Ates, T.~Pohl, S.~Sevincli, N.~Henkel, C.~Ates
  and T.~Pohl, \emph{{Nonlocal nonlinear optics in cold rydberg gases}}, Phys.
  Rev. Lett., \textbf{107}, 153001 (2011).

\bibitem{Gorshkov2011}
A.~V. Gorshkov, J.~Otterbach, M.~Fleischhauer, T.~Pohl and M.~D. Lukin,
  \emph{{Photon-Photon Interactions via Rydberg Blockade}}, Phys. Rev. Lett.,
  \textbf{107}, 133602 (2011).

\bibitem{Petrosyan2011}
D.~Petrosyan, J.~Otterbach and M.~Fleischhauer, \emph{{Electromagnetically
  Induced Transparency with Rydberg Atoms}}, Phys. Rev. Lett., \textbf{107},
  213601 (2011).

\bibitem{Otterbach2013}
J.~Otterbach, M.~Moos, D.~Muth and M.~Fleischhauer, \emph{{Wigner
  Crystallization of Single Photons in Cold Rydberg Ensembles}}, Phys. Rev.
  Lett., \textbf{111}, 113001 (2013).

\bibitem{Gorshkov2013}
A.~V. Gorshkov, R.~Nath and T.~Pohl, \emph{{Dissipative Many-body Quantum
  Optics in Rydberg Media}}, Phys. Lev. Lett., \textbf{110}, 153601 (2013).

\bibitem{Stanojevic2013}
J.~Stanojevic, V.~Parigi, E.~Bimbard, A.~Ourjoumtsev and P.~Grangier,
  \emph{{Dispersive optical nonlinearities in a Rydberg
  electromagnetically-induced-transparency medium}}, Phys. Rev. A, \textbf{88},
  053845 (2013).

\bibitem{Liu2014}
Y.~M. Liu, D.~Yan, X.~D. Tian, C.~L. Cui and J.~H. Wu,
  \emph{{Electromagnetically induced transparency with cold Rydberg atoms:
  Superatom model beyond the weak-probe approximation}}, Phys. Rev. A,
  \textbf{89}, 033839 (2014).

\bibitem{He2014a}
B.~He, A.~V. Sharypov, J.~Sheng, C.~Simon and M.~Xiao, \emph{{Two-Photon
  Dynamics in Coherent Rydberg Atomic Ensemble}}, Phys. Rev. Lett.,
  \textbf{112}, 133606 (2014).

\bibitem{Grankin2014}
A.~Grankin, E.~Brion, E.~Bimbard, R.~Boddeda, I.~Usmani, A.~Ourjoumtsev and
  P.~Grangier, \emph{{Quantum statistics of light transmitted through an
  intracavity Rydberg medium}}, New J. Phys., \textbf{16}, 043020 (2014).

\bibitem{Li2014}
W.~Li, D.~Viscor, S.~Hofferberth and I.~Lesanovsky, \emph{{Electromagnetically
  induced transparency in an entangled medium}}, Phys. Rev. Lett.,
  \textbf{112}, 243601 (2014).

\bibitem{Wu2014}
H.~Wu, M.-m. Bian, L.-t. Shen, R.-x. Chen, Z.-b. Yang and S.-b. Zheng,
  \emph{{Electromagnetically induced transparency with controlled van der Waals
  interaction}}, ArXiv:1403.5724, \textbf{1} (2014).

\bibitem{Bienias2014}
P.~Bienias, S.~Choi, O.~Firstenberg, M.~F. Maghrebi, M.~Gullans, M.~D. Lukin,
  A.~V. Gorshkov and H.~P. B{\"{u}}chler, \emph{{Scattering resonances and
  bound states for strongly interacting Rydberg polaritons}}, Phys. Rev. A,
  \textbf{90}, 053804 (2014).

\bibitem{Sommer2015}
A.~Sommer, H.~P. B{\"{u}}chler and J.~Simon, \emph{{Quantum Crystals and
  Laughlin Droplets of Cavity Rydberg Polaritons}}, ArXiv:1506.00341 (2015).

\bibitem{Grankin2015}
A.~Grankin, E.~Brion, E.~Bimbard, R.~Boddeda, I.~Usmani, A.~Ourjoumtsev and
  P.~Grangier, \emph{{Quantum-optical nonlinearities induced by Rydberg-Rydberg
  interactions: A perturbative approach}}, Phys. Rev. A, \textbf{92}, 043841
  (2015).

\bibitem{Lin2015}
G.~W. Lin, Y.~H. Qi, X.~M. Lin, Y.~P. Niu and S.~Q. Gong, \emph{{Strong photon
  blockade with intracavity electromagnetically induced transparency in a
  blockaded Rydberg ensemble}}, Phys. Rev. A, \textbf{92}, 043842 (2015).

\bibitem{Moos2015}
M.~Moos, M.~H{\"{o}}ning, R.~Unanyan and M.~Fleischhauer, \emph{{Many-body
  physics of Rydberg dark-state polaritons in the strongly interacting
  regime}}, Phys. Rev. A, \textbf{92}, 053846 (2015).

\bibitem{Maghrebi2015}
M.~F. Maghrebi, N.~Y. Yao, M.~Hafezi, T.~Pohl, O.~Firstenberg and A.~V.
  Gorshkov, \emph{{Fractional quantum Hall states of Rydberg polaritons}},
  Phys. Rev. A, \textbf{91}, 033838 (2015).

\bibitem{Maghrebi2015c}
M.~F. Maghrebi, M.~J. Gullans, P.~Bienias, S.~Choi, I.~Martin, O.~Firstenberg,
  M.~D. Lukin, H.~P. B{\"{u}}chler and A.~V. Gorshkov, \emph{{Coulomb bound
  states of strongly interacting photons}}, Phys. Rev. Lett., \textbf{115},
  123601 (2015).

\bibitem{Caneva2015}
T.~Caneva, M.~T. Manzoni, T.~Shi, J.~S. Douglas, J.~I. Cirac and D.~E. Chang,
  \emph{{Quantum dynamics of propagating photons with strong interactions: a
  generalized input-output formalism}}, New J. Phys., \textbf{17}, 113001
  (2015).

\bibitem{Shi2015}
T.~Shi, D.~E. Chang and J.~I. Cirac, \emph{{Multiphoton-scattering theory and
  generalized master equations}}, Phys. Rev. A, \textbf{92}, 053834 (2015).

\bibitem{Jachymski2016}
K.~Jachymski, P.~Bienias and H.~P. B{\"{u}}chler, \emph{{Three-body interaction
  of Rydberg slow light polaritons}}, Phys. Lev. Lett., \textbf{117}, 053601
  (2016).

\bibitem{Gullans2016}
M.~J. Gullans, J.~D. Thompson, Y.~Wang, Q.-Y. Liang, V.~Vuletic, M.~D. Lukin,
  A.~V. Gorshkov, V.~Vuleti{\'{c}}, M.~D. Lukin and A.~V. Gorshkov,
  \emph{{Effective Field Theory for Rydberg Polaritons}}, Phys. Rev. Lett.,
  \textbf{117}, 113601 (2016).

\bibitem{Murray2016b}
C.~R. Murray, A.~V. Gorshkov and T.~Pohl, \emph{{Many-body decoherence dynamics
  and optimised operation of a single-photon switch}}, New J. Phys.,
  \textbf{18}, 17 (2016).

\bibitem{Zeuthen2016}
E.~Zeuthen, M.~J. Gullans, M.~F. Maghrebi and A.~V. Gorshkov, \emph{{Correlated
  photon dynamics in dissipative Rydberg media}}, ArXiv:1608.06068 (2016).

\bibitem{Li2015a}
W.~Li and I.~Lesanovsky, \emph{{Coherence in a cold-atom photon switch}}, Phys.
  Rev. A, \textbf{92}, 043828 (2015).

\bibitem{Murray2017}
C.~R. Murray and T.~Pohl, \emph{{Coherent photon manipulation in interacting
  atomic ensembles}}, Arxiv:1702.03763 (2017).

\bibitem{Bienias2016}
P.~Bienias and H.~P. B{\"{u}}chler, \emph{{Quantum theory of Kerr nonlinearity
  with Rydberg slow light polaritons}}, New J. Phys., \textbf{18}, 123026
  (2016).

\bibitem{Friedler2005}
I.~Friedler, D.~Petrosyan, M.~Fleischhauer and G.~Kurizki, \emph{{Long-range
  interactions and entanglement of slow single-photon pulses}}, Phys. Rev. A,
  \textbf{72}, 043803 (2005).

\bibitem{Paredes-Barato2014a}
D.~Paredes-Barato and C.~S. Adams, \emph{{All-optical quantum information
  processing using Rydberg gates}}, Phys. Rev. Lett., \textbf{112}, 040501
  (2014).

\bibitem{Khazali2015}
M.~Khazali, K.~Heshami and C.~Simon, \emph{{Photon-photon gate via the
  interaction between two collective Rydberg excitations}}, PRA, \textbf{91},
  030301(R) (2015).

\bibitem{Shapiro2006}
J.~Shapiro, \emph{{Single-photon Kerr nonlinearities do not help quantum
  computation}}, Phys. Rev. A, \textbf{73}, 062305 (2006).

\bibitem{Shapiro2007}
J.~H. Shapiro and M.~Razavi, \emph{{Continuous-time cross-phase modulation and
  quantum computation}}, New J. Phys., \textbf{9}, 16--16 (2007).

\bibitem{Gea-Banacloche2010}
J.~Gea-Banacloche, \emph{{Impossibility of large phase shifts via the giant
  Kerr effect with single-photon wave packets}}, Phys. Rev. A, \textbf{81},
  043823 (2010).

\bibitem{Masalas2004}
M.~Ma{\v{s}}alas and M.~Fleischhauer, \emph{{Scattering of dark-state
  polaritons in optical lattices and quantum phase gate for photons}}, Phys.
  Rev. A, \textbf{69}, 061801 (2004).

\bibitem{Andre2005}
A.~Andr{\'{e}}, M.~Bajcsy, A.~Zibrov and M.~Lukin, \emph{{Nonlinear Optics with
  Stationary Pulses of Light}}, Phys. Rev. Lett., \textbf{94}, 063902 (2005).

\bibitem{Shahmoon2011}
E.~Shahmoon, G.~Kurizki, M.~Fleischhauer and D.~Petrosyan, \emph{{Strongly
  interacting photons in hollow-core waveguides}}, Phys. Rev. A, \textbf{83},
  33806 (2011).

\bibitem{Marzlin2010}
K.~Marzlin, Z.~Wang, S.~Moiseev and B.~Sanders, \emph{{Uniform cross-phase
  modulation for nonclassical radiation pulses}}, JOSA B, \textbf{27}, 36--45
  (2010).

\bibitem{He2012}
B.~He and A.~Scherer, \emph{{Continuous-mode effects and photon-photon phase
  gate performance}}, Phys. Rev. A, \textbf{85}, 033814 (2012).

\bibitem{SuppBienias}
\emph{{See Supplementary Material for detailed derivation.}}

\bibitem{Tiarks2019}
D.~Tiarks, S.~Schmidt-Eberle, T.~Stolz, G.~Rempe and S.~D{\"{u}}rr, \emph{{A
  photon–photon quantum gate based on Rydberg interactions}}, Nat. Phys.,
  \textbf{15}, 124--126 (2019).

\bibitem{Thompson2017}
J.~D. Thompson, T.~L. Nicholson, Q.~Y. Liang, S.~H. Cantu, A.~V. Venkatramani,
  S.~Choi, I.~A. Fedorov, D.~Viscor, T.~Pohl, M.~D. Lukin and V.~Vuletic,
  \emph{{Symmetry-protected collisions between strongly interacting photons}},
  Nature, \textbf{542}, 206--209 (2017).

\bibitem{Busche2017}
H.~Busche, P.~Huillery, S.~W. Ball, T.~Ilieva, M.~P. Jones and C.~S. Adams,
  \emph{{Contactless nonlinear optics mediated by long-range Rydberg
  interactions}}, Nat. Phys., \textbf{13}, 655--658 (2017).

\bibitem{Hsiao2018}
Y.-F. Hsiao, P.-J. Tsai, H.-S. Chen, S.-X. Lin, C.-C. Hung, C.-H. Lee, Y.-H.
  Chen, Y.-F. Chen, I.~A. Yu and Y.-C. Chen, \emph{{Highly efficient coherent
  optical memory based on electromagnetically induced transparency}}, Phys.
  Rev. Lett., \textbf{120}, 183602 (2016).

\bibitem{Gorshkov2007}
A.~Gorshkov, A.~Andr{\'{e}}, M.~Lukin and A.~S{\o}rensen, \emph{{Photon storage
  in $\Lambda$-type optically dense atomic media. II. Free-space model}}, Phys.
  Rev. A, \textbf{76}, 1--25 (2007).

\bibitem{Asenjo-Garcia2017a}
A.~Asenjo-Garcia, M.~Moreno-Cardoner, A.~Albrecht, H.~J. Kimble and D.~E.
  Chang, \emph{{Exponential improvement in photon storage fidelities using
  subradiance and "selective radiance" in atomic arrays}}, Phys. Rev. X,
  \textbf{7}, 031024 (2017).

\bibitem{Bienias2016a}
P.~Bienias, \emph{{Few-body quantum physics with strongly interacting Rydberg
  polaritons}}, Eur. Phys. J. ST, \textbf{225}, 2957 (2016).

\bibitem{Gorshkov2011b}
A.~Gorshkov, J.~Otterbach, M.~Fleischhauer, T.~Pohl and M.~Lukin,
  \emph{{Photon-Photon Interactions via Rydberg Blockade}}, Phys. Rev. Lett.,
  \textbf{107}, 1--4 (2011).

\bibitem{He2011}
B.~He, A.~MacRae, Y.~Han, A.~I. Lvovsky and C.~Simon, \emph{{Transverse
  multimode effects on the performance of photon-photon gates}}, Phys. Rev. A,
  \textbf{83}, 022312 (2011).

\bibitem{Lloyd1995}
S.~Lloyd, \emph{{Almost any Quantum Gate is universal}}, Phys. Rev. Lett.,
  \textbf{75}, 346 (1995).

\bibitem{Rauschenbeutel1999}
A.~Rauschenbeutel, G.~Nogues, S.~Osnaghi, P.~Bertet, M.~Brune, J.~Raimond and
  S.~Haroche, \emph{{Coherent Operation of a Tunable Quantum Phase Gate in
  Cavity QED}}, Phys. Rev. Lett., \textbf{83}, 5166--5169 (1999).

\end{thebibliography}
\bibliographystyle{bernd}

\clearpage
\newpage
\beginsupplement
\setcounter{equation}{0}
\renewcommand{\theequation}{S\arabic{equation}}
\appendix

\ioptwocol
\onecolumn
\section{Corrections due to inhomogeneous distribution of atoms.}
Analogously to \cite{Bienias2016} we can include a shape of atomic distribution in the solution  of the two-body counterpropagating problem. 
In this section, we will use the notation from \cite{Bienias2016}, i.e., $\beta(x)$ describes the amplitude of the polariton to
be in a photonic state and is related to the slow light velocity $v_{g} = c \beta(x)^2$, while $n(x) = 1- \beta(x)^2$ is
the probability for the polariton to be in the Rydberg state. These quantities are determined by 
the atomic density $n_{\rs at}(x)$ via $\beta(x) = {\Omega}/{\sqrt{\Omega^2 + g_{0}^2 n_{\rs at}(x) }}$ 
with $g_{0}$ the single atom coupling. Then, the solution  of two-body problem
\begin{equation}
\phi^{\rs out}(x,y,t) =e^{- i \varphi(x,y)}     \phi^{\rs in}(x-c t' ,y+c t')
\label{twoPhoton}
\end{equation}
where $t' = t-\Delta t$  accounts for the delay of the polaritons inside the medium with $\Delta t = \int_{-\infty}^{\infty} \textrm{d}y  \left({1}/{ \beta(y)^{2}}-1\right)/c$. 
The phase factor  $ \varphi $
describes the  correlations built up between the photons during the propagation through the medium and takes the form
\begin{eqnarray}
 \varphi(z,z',t)	&	=	&	
													\frac{1}{\hbar c}	\integralb{z-ct}{z}{w}%
													\tilde{n}(w) \tilde{n}(z+z'-w) 	\tilde{V}(z+z'-w,w) 
\end{eqnarray}
where $\tilde{n}(z) = n(\zeta(z))$, $\tilde{V}(z,w) = V(\zeta(z)- \zeta{(w)})$ with the coordinate transformation taking the form 
$ z=\zeta^{-1}(x) =\int_{0}^{x}\textrm{d}y \left({1}/{ \beta(y)^{2}}\right)$.
For large $t$ and $z$ this integral is equivalent to the integral from $-\infty$ to $\infty$.
For Gaussian distribution of atoms much longer than the blockade radius, $\mathcal{L}\gg\xi$, the phase factor $\varphi$ simplifies to
\begin{eqnarray}
\varphi(z,z')&=&\frac{1}{\hbar c}\integralb{}{}{w} \nt\left(\frac{z+z'}{2}\right)^2 V(\zeta(z+z'-w)-\zeta(w))
\end{eqnarray}
and can be furthermore transformed using $u=\frac{z+z'}{2}$ and  $y=\zeta(u)$,
\begin{eqnarray}
\varphi(z,z')&=&\frac{1}{\hbar c}\nt\left(u\right)^2\integralb{}{}{w}  V(\partial_{u}\zeta(u) 2w)\\
\varphi(z,z')&=&\frac{1}{\hbar c}\nt(u)^2\integralb{}{}{w}  V(\beta(y)^2 2w)
\end{eqnarray}
The integral can be calculated analytically, leading to
\beqa
\varphi(z,z')&=&\frac{1}{\hbar c} n(y)^2\frac{2\pi}{3} \frac{\xi}{2 \beta(y)^2}\frac{2\Omega^2}{\Delta} %
\eeqa
For $g\gg\Om$ and compressed pulses shorter than the cloud's length, we can set $n(y)=1$, leading to
\beqa
\varphi(z,z')&\approx&\frac{1}{\hbar c}\frac{2\pi}{3} \frac{g_0^2 n_{at}(y)}{\Omega^2}\frac{\xi}{2 }\frac{2\Omega^2}{\Delta}
\eeqa
For small $y$, using that $u=\zeta^{-1}(y)=\frac{g_0^2n_0}{\Om^2}\frac{\sqrt{\pi}}{2}\mathcal{L}\,\textrm{Erf}(y/\mathcal{L})+y\approx \left(1+\frac{g_0^2n_0}{\Om^2}\right)y $, we can write 
\begin{equation}
y=\zeta\left(u\right)\approx \left(1+\frac{g_0^2n_0}{\Om^2}\right)^{-1}u,
\end{equation}
which gives
\begin{eqnarray}
\varphi\left(u=\frac{z+z'}{2}\right)=\varphi_0 \exp\left[-\left(\frac{u v_g}{c \mathcal{L}}\right)^2\right],
\end{eqnarray}
from which,
\begin{eqnarray*}
\beta_{\rs at}e^{i\phi_0+i\Delta\phi_{\rs at}}
=\lim_{t\ra \infty}\braket{\mathcal{E}\mathcal{E}}{\mathcal{E}\mathcal{E}^V}=\lim_{t\ra \infty}\integral{z}\textrm{d}z' \,e^{i\phi_0} \mathcal{N}^2 e^{-(z-z_0-ct)^2/s_z^2}e^{-(z'+z_0+ct)^2/s_z^2} 
\exp\left[
i\phi_0\left(e^{-\left(\frac{u v_g}{c \mathcal{L}}\right)^2}-1\right)
\right]\,,\nn\\
\end{eqnarray*}
where as initial conditions we have chosen two Gaussian wavepackets centered at positions $\pm z_0$ with width $s_z$ and 
 normalization factor $\mathcal{N}$.
This integral can be calculated analytically for $s_z/c\ll \mathcal{L}/v_g $, leading to
\begin{equation}
\beta_{\rs at}e^{i\Delta\phi_{\rs at}}= \left(1+\frac{i \phi _0 v_g^2 s_z^2}{2 c^2 \mathcal{L}^2}\right)^{-1/2} \,
.
\end{equation}
Expanding it in small parameter $\frac{ s_z v_g}{ c \mathcal{L}}$ we get
\begin{eqnarray}
\beta_{\rs at}
&=&1 -\frac{3}{2}\phi _0^2\left(\frac{ s_z   v_g}{2 c \mathcal{L}}\right)^4\,, \\
\D\phi_{\rs at}&=&-\phi _0\left(\frac{ s_z   v_g}{2 c \mathcal{L}}\right)^2\,,
\end{eqnarray}
which corresponds to the expressions from the main text because  $s_z=\sigma_z c/v_g$.

\section{Transversal size effects}
The propagation of photons within paraxial approximation is described by the  Hamiltonian (analogous to Eq.~\ref{quadraticHamiltonian-1D}) given by 
\begin{eqnarray}
 H_{\pm} = \hbar \integral {\bz} \left(
     \begin{array}{c}
      	    \psie_\pm(\bz)\\
	    \psip_\pm(\bz)\\
	    \psis_\pm(\bz)
     \end{array}\right)^{\dag}
     \left(\begin{array}{ccc}
   \mp i  c \partial_{z}-\frac{c}{2k_p}\nabla_\perp^2& g_0\sqrt{n(\bz)}    &  0 \\
         g_0\sqrt{n(\bz)}    &   \Delta &  \Omega\\
        0  &  \Omega & 0
    \end{array} \right)
        \left(
     \begin{array}{c}
      	    \psie_\pm(\bz)\\
	    \psip_\pm(\bz)\\
	    \psis_\pm(\bz)
     \end{array}\right).
     \label{eq:quadraticHamiltonian-quasi1D}
\end{eqnarray}
where $k_p$ is the carrier frequency of the probe photons.
We are interested in an estimate of leading corrections to the fidelity, thus 
we neglect transversal corrections to the dispersion relation of photons.%
In addition, we consider atomic distribution broader than the extend of relevant photonic modes, leading to
\begin{equation}
 H_{\pm} = \hbar \integral {\bz} \left(
     \begin{array}{c}
      	    \psie_\pm(\bz)\\
	    \psip_\pm(\bz)\\
	    \psis_\pm(\bz)
     \end{array}\right)^{\dag}
     \left(\begin{array}{ccc}
   \mp i  c \partial_{z}& g_0\sqrt{n_{\rs at}}    &  0 \\
         g_0\sqrt{n_{\rs at}}    &   \Delta &  \Omega\\
        0  &  \Omega & 0
    \end{array} \right)
        \left(
     \begin{array}{c}
      	    \psie_\pm(\bz)\\
	    \psip_\pm(\bz)\\
	    \psis_\pm(\bz)
     \end{array}\right).
     \label{quadraticHamiltonian-quasi1Db}
\end{equation}
Then, the solution of interacting two-body problem can be derived analogously to Eq.~\ref{eq:TMatrix1D}
leading to
\begin{eqnarray}
   \psi_{k_{\rs D}}^{ss}(r,\bR_+,\bR_-) &= & \frac{1}{1-\bar{\chi} V(\br)}u_{00}(\bR_+)u_{00}(\bR_-)e^{i k_{\rs D} r} \exp\left[ - i \frac{\alpha_{\rs D}}{\hbar} \int_{-\infty}^{r} \textrm{d}z V^{\rs eff}(|\bR_+-\bR_-|,z)\right]  
\end{eqnarray}
where $\br=\{r,\bR=\bR_+-\bR_-\}$ with $r$ being the longitudinal component of relative distance, and $\bR_\pm$ are transverse components of the $\bz_\pm$ ($\bz_\pm$ corresponds  to $\bz$ in \eqref{quadraticHamiltonian-quasi1Db} for two directions of the propagation).
$u_{00}$ is the lowest 
 Laguerre-Gauss mode describing transverse shape of the incoming photons (in the main text $u_{00}$ is denoted by $u$).  
 This solution in the limit of $r\ra \infty$ leads to the expression for $\beta_{\rs tr}$ from the main text.

\section{%
Summary of the corrections to the fidelity}
\begin{table*}[h!]
\center
\begin{tabular}{c c  }
$\beta_{j}$& estimate  \\% \hspace{30pt} 
 \hline
\\[-0.2cm]
$\beta_{\gamma}$& $e^{- 2 \eta}$, with  $\eta=\phi \frac{5}{6}\frac{\gamma}{\delta}$    \\
\\[-0.3cm]
$|\beta_{\rs sc}|$& $   1-  \phi^2\frac{\Omega^2}{|\Delta|^2}\frac{|\Delta|^2+g^2}{4g^2}\,,
$    \\
\\[-0.3cm]
$\beta_{\rs wp}$&\hspace{2em} $P_{\rs +} P_{\rs -}  \exp \left[ -
	\frac{49}{2}\phi^2  \left( \frac{\sigma |\Delta|}{12 \Omega^2} \right)^2  \right]$, where $P_{\rs \pm} =  \exp \left[ -2 \frac{Lg^2}{\delta c} \left(\frac{\delta \sigma}{\Om^2}\right)^2\frac{\gamma}{\delta}
 \right] $    \\
 \\[-0.3cm]
$\beta_{\rs at}$& $1  -\frac{3}{2}\phi ^2\left(\frac{ \sigma_z}{2 \mathcal{L}}\right)^4 \,.
 $    \\
 \\[-0.3cm]
$\beta_{\rs tr}$& $\mathcal{F}(\sRp/\xi)
 $    \\
  \\[-0.3cm]
$\beta_{\rs Ry}$& $e^{-4\gamma_{s}L/v_g}
 $    \\
\end{tabular}
\label{table1}
\caption{ \label{table1} Summarizing table of analyzed contributions to the fidelity.} %
\end{table*}
In the TAB.~\ref{table1} we sum-up all the estimates for the corrections to the phase gate fidelity $F$, given by \eqref{eq:FfromBeta}. 
In order to generate results presented in \figref{fig:estimates} and \figref{fig:estimates2}, we maximize the value of $\sqrt{F}$ 
for fixed $\sRp$, $\phi, C_6$  in function of $g,L$, by finding optimal detuning $\delta$. Then, the value of Rabi frequency $\Om$ is fixed by the constraint on $\phi$.
 \section{Results for alternative definition of fidelity: conditional-fidelity}
\begin{figure}[h!]
\center
\includegraphics[width= 0.45\columnwidth]{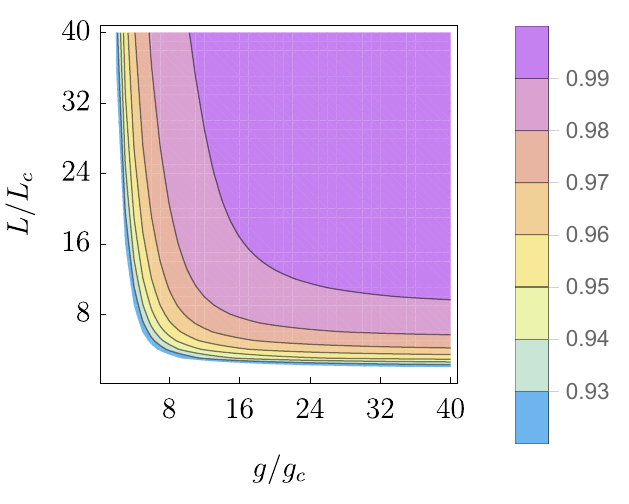}
\caption{
The $\sqrt{F_{\rs cond}}$ which contains all presented corrections except single body losses (i.e. we set $P_\pm=1$ and $\beta_{\rs Ry}=1$)  for parameters as in \figref{fig:estimates}. %
}
\label{fig:estimates2}
\end{figure}

 In \figref{fig:estimates2} we show the conditional-fidelity $\sqrt{F_{\rs cond}}$ which contains all presented corrections except single body losses (i.e. we set $P_\pm=1$ and $\beta_{\rs Ry}=0$)  for parameters as in \figref{fig:estimates}.
 \newpage
 \textcolor{white}{nothing}
\clearpage

 \end{document}